\documentclass[10pt,twocolumn]{IEEEtran}
\usepackage{amssymb, amsmath, graphicx, cite, epsfig, booktabs}

\usepackage[usenames,dvipsnames]{pstricks}
\usepackage{epsfig}
\usepackage{pst-grad} 
\usepackage{pst-plot} 

\newcommand{\posreals}{\mathbb{R}_{+}}

\newcommand{\Fig}   {\mbox{Fig.} }

\newcommand{\eqdef}{ := }

\newcommand{\bc}{ {\bf c} }
\newcommand{\bd}{ {\bf d} }

\newcommand{\snr} { {\sf {snr}} }

\newcommand{\pr} { {\sf {Pr}} }

\newcommand{\D}{ {\Delta} }
\newcommand{\Xa}[2]{{X_{a_#1}^{(#2)}}}
\newcommand{\Xr}{X_{r}^{(2)}}
\newcommand{\Ya}[2]{{Y_{a_#1}^{(#2)}}}
\newcommand{\Yb}[2]{{Y_{b_#1}^{(#2)}}}
\newcommand{\Yr}{Y_{r}^{(1)}}

\newcommand{\Zb}[2]{{Z_{b_#1}^{(#2)}}}

\newcommand{\Zr}{Z_{r}^{(1)}}
\newcommand{\bZr}{{\mathbf{Z}}_{r}^{(1)}}
\newcommand{\bXa}[2]{{{\mathbf{X}}_{a_#1}^{(#2)}}}
\newcommand{\bXr}{{\mathbf{X}}_{r}^{(2)}}

\newcommand{\bYb}[2]{{{\mathbf{Y}}_{b_#1}^{(#2)}}}
\newcommand{\bYr}{{\mathbf{Y}}_{r}^{(1)}}
\newcommand{\bX}[2]{{{\mathbf{X}}_{#1}^{(#2)}}}
\newcommand{\bY}[2]{{{\mathbf{Y}}_{#1}^{(#2)}}}

\newtheorem{theorem}{Theorem}

\newtheorem{corollary}{Corollary}

\newcommand{\qed}{\nobreak \ifvmode \relax \else
      \ifdim\lastskip<1.5em \hskip-\lastskip
      \hskip1.5em plus0em minus0.5em \fi \nobreak
      \vrule height0.35em width0.4em depth0.15em\fi}

\newenvironment{proof1}{\noindent{\em Proof Outline}:}
	{\hfill\qed\vspace{1ex}}
\renewenvironment{proof}{\noindent{\em Proof}:}
	{\hfill\qed\vspace{1ex}}

\allowdisplaybreaks[2]
\renewcommand{\baselinestretch}{0.94}

\begin{document}

\title{\vspace{-0.8cm}Capacity Bounds and Lattice Coding for the Star Relay Network}
\author{{H.~Ebrahimzadeh Saffar and P.~Mitran \\
Department of Electrical and Computer Engineering \\
University of Waterloo, Waterloo, Ontario, Canada\\
Email: \tt\{hamid, pmitran\}@ece.uwaterloo.ca} \vspace{-0.99cm}}



\maketitle

\begin{abstract}
A half-duplex wireless network with $6$ lateral nodes, $3$ transmitters and $3$ receivers, and a central relay is considered.
The transmitters wish to send information to their corresponding receivers via a two phase communication
protocol. 
The receivers decode their desired messages by using side information and the signals received from the relay. We derive an
outer bound on the capacity region of any two phase protocol as well as $3$ achievable regions by employing different relaying
strategies. In particular, we combine physical and network layer coding to take advantage of the interference at the relay,
using, for example, lattice-based codes. We then specialize our results to the exchange rate. It is shown that for any $\snr$,
we can achieve within 
$0.5$ bit of the upper bound by lattice coding and within $0.34$ bit, if we take the best of the $3$ strategies. Also, for high
$\snr$, lattice coding is within $\log(3)/4 \simeq 0.4$ bit of the upper bound.

\end{abstract}


\section{Introduction}
\label{sec:Introduction} Relaying in wireless communications has emerged to be a major subject of research in network
information theory. In particular, bidirectional communication between two terminal nodes $a$, and $b$, with the aid of a relay
node $r$ has been of interest to improve communication quality \cite{Katti:2006}, \cite{Popovski:2006a}, \cite{Wu:2005}. These
works use network coding at the relay to increase the information exchange rate between $a$ and $b$ by XORing the packets from
$a$ and $b$ at the relay and broadcasting the result to the terminal nodes. An exception is \cite{Katti:2007}, where a natural
combination of signals at the physical layer is considered.

Unlike \cite{Katti:2007}, where no coding is performed at the relay node, some recent works perform coding at the physical
layer by using structured codes and in particular codes defined on lattices \cite{Narayanan:2007, Nazer_Gastpar:2009},
which can be considered as joint physical layer and network coding. A common aspect of these works is the exploitation of
interference at the relay to obtain higher rates, based on either an amplify-and-forward (AF) relaying strategy
\cite{Katti:2007}, or using lattice-based codes, \cite{Narayanan:2007}, \cite{Nazer_Gastpar:2009}. For codes with group
structure (e.g., lattice-based codes), the modular sum
of the codewords are themselves codewords, 
thus the decoder should search a single codebook, as opposed to a product codebook, resulting in higher rates.

Besides 
relaying and the use of cross layer coding, there has also been interest in taking advantage of side information in the
wireless environments to improve performance \cite{Katti_Katabi_Hu:2005, Effros_Ho_Kim:2006}. In practical wireless networks
with a large number of users (e.g., mesh networks), overhearing nodes (opportunistic listening, \cite{Katti_Katabi_Hu:2005}) in
different phases of a communication protocol becomes vital. However, exploiting side information for lattice coding is not
addressed in the literature.
\begin{figure}[ht]
{\centering{\includegraphics*[angle = 90,viewport=-154 388 -66 608]
{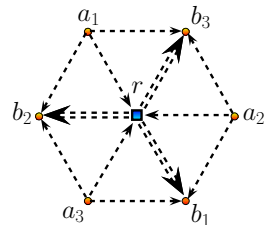}} }
\caption{Star relay network with the two phase (MBC)
protocol. Single and double arrows show the transmissions in phases 1 and 2, respectively.} \vspace{-.6cm}\label{System}
\end{figure}

Motivated by these observations, we consider a problem in which three transmitter terminals desire to unicast information to
three receivers by means of a central relay in two consecutive phases. In phase 1, the transmitters broadcast the encoded
messages. The terminals are located on vertices of a hexagon while the half-duplex relay is at the center and each node can
only hear its immediate neighbor nodes. This network, referred to as the {\em star network}, is a constructive component of the
triangular lattice with error free links addressed in \cite{Effros_Ho_Kim:2006}, where the network code design is based on the
fact that each node receives side information from its neighbors at the end of the first phase. Although we consider the use of
phase 1 side information, unlike \cite{Effros_Ho_Kim:2006}, we assume non-ideal channels and also let the central relay perform
joint physical and network layer coding. In phase 2, the relay broadcasts the result to the receivers and finally, each
receiver extracts its own desired message, using the side information from the first phase and reception from the second phase.

The contributions of this paper are as follows. We derive an outer bound on the capacity region and also achievable rate
regions for three different communication strategies, namely, decode-and-forward (DF), amplify-and-forward (AF), and lattice
coding, of the star network. We then examine the exchange rate and compare the performance of the three strategies.
\section{Preliminaries}
\label{sec:preliminaries}
\subsection{Network Model}
We consider a wireless half-duplex communication network with six terminal nodes that are located around a central relay. The
terminal nodes are divided into three transmitter (source) and receiver (sink) pairs $\{a_i,b_i\}_{i=1}^{3}$. Each transmitter
node $a_i$ requires to send information with rate $R_i$ to the corresponding receiver $b_i$ with the aid of the relay. In case
where $R=R_1=R_2=R_3$, we call $R$ the {\em exchange rate}. We assume that each node can overhear the signals only from its
immediate neighbors. It is also assumed that the network uses a two-phase communication protocol. In the first phase, the
lateral sources transmit their messages simultaneously and in the second phase, the relay broadcasts some function of the
superposed signals received in the first phase. It is sometimes relevant to call phases 1 and 2 of the protocol multiple access
channel (MAC) and broadcast (BC) phases respectively, and thus refer to the communication protocol as MBC. The sink nodes have
to estimate their desired messages by use of the signals they received after both phases. \Fig \ref{System} depicts a symmetric
schematic of the network, transmissions, and the relative position of the terminal nodes and the relay. The lateral nodes and
the relay form a hexagon with the length of the edges being less than the transmission ranges. The topology of the system,
thus, is not necessarily symmetric.

For the sake of practical comparison between different strategies and design of lattice codes, we consider the Gaussian case in
this paper. In this scenario, all of the channels in the system are assumed to be additive white Gaussian noise (AWGN). 
\vspace{-0.31cm}
\subsection{Notations and Definitions}
For the communication network shown in \Fig \ref{System} and every node $i \in \left\{a_1,b_1,a_2,b_2,a_3,b_3,r\right\}$, we
denote the channel input and output signals in phase $m$ by random variables $X_{i}^{(m)}$ and $Y_{i}^{(m)}$, respectively.
$X_{i}^{(m)}$ is chosen from alphabet $\mathcal{X}_{i}$ and input distribution $p^{(m)}(x_{i})$. Boldface vectors $\bX{i}{m}$
and $\bY{i}{m}$ represent the sent or received vectors of node $i$, respectively, indexed by phase number $m$. The message that
node $i$ wishes to send to the corresponding receiver is denoted by $w_i$. It is also convenient to denote the additive white
Gaussian noise at node $j$, in phase $m$, by $Z_{j}^{(m)}$ for a given time and by ${\bf{Z}}_{j}^{(m)}$ for a vector reception.
Hence, for the Gaussian case, we have {
\begin{align}
\hspace{-.48cm}\Yb{i}{1} = &\Xa{j}{1} + \Xa{k}{1} +  \Zb{i}{1}, (i,j,k) \in {\mathbb{P}_{3}}, \label{Gaussian1}\\
\Yr & = \Xa{1}{1} + \Xa{2}{1} + \Xa{3}{1} + \Zr\label{Gaussian2}\\
\Yb{i}{2} & = \Xr + \Zb{i}{2}, i = 1, 2, 3, \label{Gaussian3}
\end{align}} \noindent \hspace{-0.23cm} where $\mathbb{P}_{3} \triangleq \{(1,2,3),(2,3,1),(3,1,2)\}$, and all of the nodes have the same power constraint
$P$ and the complex circular symmetric noise has variance $N$. The signal-to-noise ratio $P \over N$ is denoted by $\snr$.
Furthermore, we denote the set of transmissions and receptions by all nodes in a set $S$ at a given time, in phase $m$, by
$X_{S}^{(m)}$ and $Y_{S}^{(m)}$ respectively. We also define $R_{S} \triangleq \sum_{i\in S}R_{i}$.

For a given communication strategy, $\Delta_{m} \geq 0$, denotes the relative duration of the $m$th phase of the protocol and
we have $\sum_{m}\D_{m} = 1$. In particular, for the present MBC protocol, $\D_1 + \D_2 = 1$. For a given phase $m$, the error
events $\{\hat{w}_{a_{i}} \neq {w_{a_i}}\}$ between encoder $a_i$ and decoders are denoted by $E_{ij}^{(m)}$, if $b_j$ is the
decoder, and by $E_{ir}^{(m)}$, if the relay is the decoder. Finally, we denote $\epsilon$-weakly typical sequences of length
$n\D_m$, according to distributions of phase $m$, by $T_{\epsilon}^{(m)}$. \vspace{-0.21 cm}
\section{Outer Bound}
To derive the outer bound on the three rates of the star network with the MBC protocol, we use the variation of the cut-set
bound given in \cite{SKim:2008}. By looking at all cut-sets and dropping those implied by the other ones, we find six upper
bounds on each of the single rates $R_1$, $R_2$, and $R_3$, three upper bounds for two-term sums, and two upper bounds for the
sum-rate.

The cut-sets that give the upper bounds on $R_1$ are
${\left\{b_1\right\}}^c$, ${\left\{a_3,b_1\right\}}^c $,
${\left\{a_2,b_1\right\}}^c$, ${\left\{a_2,a_3,b_1\right\}}^c$,
${\left\{a_1,b_2,b_3\right\}}$ and ${\left\{a_1,r\right\}}$. Based
on these respectively, $R_1 \leq \bar{\mathcal{C}}_1$, where
\begin{align}
\overline{\mathcal{C}}_1 = \min \Bigl\{ \Bigr. \D_1
I(X_{a_1}^{(1)},X_{a_2}^{(1)},X_{a_3}^{(1)}
; &Y_{b_1}^{(1)}) +  \D_2 I(X^{(2)}_{r};Y^{(2)}_{b_1}),  \nonumber \\
\D_1 I(\Xa{1}{1},\Xa{2}{1};Y_{b_1}^{(1)}\vert \Xa{3}{1})  + \D_2 &I(\Xr;\Ya{3}{2},\Yb{1}{2}),  \nonumber \\
\D_1 I(\Xa{1}{1},\Xa{3}{1};Y_{b_1}^{(1)}\vert \Xa{2}{1})  + \D_2 &I(\Xr;\Ya{2}{2},\Yb{1}{2}), \nonumber
\end{align}
\begin{align}
\D_2 I(\Xr;\Ya{2}{2},\Ya{3}{2},\Yb{1}{2}), \ \D_1  I&(\Xa{1}{1},\Yr \vert \Xa{2}{1},\Xa{3}{1}), \nonumber \\
\Bigl. \D_1 I(\Xa{1}{1};\Yb{2}{1},\Yb{3}{1} \vert
\Xa{2}{1},\Xa{3}{1}) &+  \D_2 I(\Xr;Y_{{\{a_1,r\}}^{c}}^{(2)})
\Bigr\}. \nonumber
\end{align}\par
There are six similar upper bounds on $R_2$ and $R_3$, derived from corresponding cut-sets. The minimum of the six bounds on
$R_2$ and $R_3$ are denoted by ${\overline{\mathcal{C}}}_{2}$ and $\overline{\mathcal{C}}_3$, respectively.

We also bound the two-term sums such as $R_1+R_2$ by making appropriate cut-sets. By using cut-sets
${\left\{a_1,a_2,b_3,r\right\}}$, ${\left\{b_1,b_2\right\}}^c$, and ${\left\{a_1,a_2,b_3\right\}}^c$, respectively, we derive
the upper bound $R_1 + R_2 \leq \overline{\mathcal{C}}_{1,2}$, where{\small{
\begin{align}
\overline{\mathcal{C}}_{1,2} & = \min \Bigl\{ \Bigr. \D_1
I(X_{a_1}^{(1)},X_{a_2}^{(1)};
Y_{b_1}^{(1)},\Yb{2}{1} \vert \Xa{3}{1}) \ + \nonumber \\
\D_2 & I(X^{(2)}_{r};Y^{(2)}_{b_1}, \Yb{2}{2},\Ya{3}{2}),
\D_1 I(\Xa{1}{1},\Xa{2}{1},\Xa{3}{1};Y_{b_1}^{(1)},\Yb{2}{1}) + \nonumber \\
\D_2 & I(\Xr;\Ya{1}{2}, \Yb{2}{2}), \D_1
I(\Xa{1}{1},\Xa{2}{1};\Yr,\Yb{1}{1},\Yb{2}{1} \vert \Xa{3}{1})
\Bigl. \Bigr\}.\nonumber
\end{align}}}
By using symmetric cut-sets, similar bounds are found on $R_2+R_3$ (denoted by $\overline{\mathcal{C}}_{2,3}$) and $R_1+R_3$
(denoted by $\overline{\mathcal{C}}_{1,3}$).

Finally, by considering the cut-sets ${\left\{a_1,a_2,a_3,r\right\}}$ and ${\left\{a_1,a_2,a_3\right\}}^c$, we derive two upper
bounds on the sum-rate $R_1+R_2+R_3$ and we choose the tighter one as the ultimate upper bound on the sum-rate which can be
written as
\begin{align}
\overline{\mathcal{C}}_{1,2,3}
 & = \min \Bigl\{ \Bigr. \D_1 I(X_{a_1}^{(1)},X_{a_2}^{(1)},\Xa{3}{1};
Y_{b_1}^{(1)},\Yb{2}{1},\Yb{3}{1})\nonumber \\
& + \D_2 I(X^{(2)}_{r};Y^{(2)}_{b_1}, \Yb{2}{2},\Ya{3}{2}), \nonumber \\
& \left. \D_1
I(\Xa{1}{1},\Xa{2}{1},\Xa{3}{1};\Yr,Y_{b_1}^{(1)},\Yb{2}{1},\Yb{3}{1})
\right\}. \nonumber
\end{align}
Therefore, an outer bound on the capacity region of this network, denoted by $\overline{\mathcal{C}}$ is described by
\begin{align}
\overline{\mathcal{C}} = & \Bigl\{(R_1,R_2,R_3) \in \posreals^{3}:
\Bigr. \ R_1 \leq {\overline{\mathcal{C}}}_{1}, \ R_2 \leq
{\overline{\mathcal{C}}}_{2}, \ R_3 \leq {\overline{\mathcal{C}}}_{3} \nonumber \\
& R_1+R_2 \leq \overline{\mathcal{C}}_{1,2}, \ R_2+R_3 \leq
\overline{\mathcal{C}}_{2,3}, \ R_1+R_3 \leq
\overline{\mathcal{C}}_{1,3}\nonumber \\
& \Bigl. R_1+R_2+R_3 \leq \overline{\mathcal{C}}_{1,2,3} \Bigr\}. \nonumber
\end{align}
{\em Gaussian case}: Computing the outer region $\overline{\mathcal{C}}$ for the Gaussian case and dropping degenerate
inequalities, we have that the exchange rate $R$ is bounded by
\begin{align}
& R \leq \min \Bigl( \D_1 \log(1+\snr), \D_2 \log(1+3\snr)\Bigr), \label{upper1}\\
& 2R \leq \min \Bigl(\D_1 \log(1+4\snr+3\snr^2), \Bigr. \nonumber \label{upper2}\\
& \qquad \qquad \Bigl. \D_1\log(1+\snr)^2 + \D_2 \log(1+3\snr) \Bigr),   \\
& 3R \leq  \min \Bigl( \D_1 \log\left[(1+\snr)^2(1+7\snr)\right] \Bigr.,
\nonumber \\
& \Bigl. \D_1 \log\left[{(1+\snr)^2(1+4\snr)}\right] +
\D_2\log(1+3\snr) \Bigr), \label{upper3}
\end{align}
\noindent where all logarithms are in base $2$. It is easy to see that (\ref{upper1}) implies (\ref{upper2}) and
(\ref{upper3}), thus (\ref{upper1}) represents the tightest cut-set upper bound on the exchange rate of the star network for
the Gaussian case within a two-phase protocol.
\section{Listening and Relaying Strategies}
\subsection{Decode-and-Forward (DF)} In the 
decode-and-forward protocol, the relay tries to decode the messages sent by the transmitters separately and then forwards a
combination of them to the receivers. Each receiver node also decodes the messages of the other two pairs at the end of the
first phase, using the side information received during the first phase. Consequently, at the end of the second phase, each
receiver has a combination of all three messages as well as the two messages of other two pairs. Thus, the receivers can
extract the desired messages.
\begin{theorem}
\label{Theorem:DF} An achievable region for the star relay network with the two phase DF MBC protocol is the closure of the
convex hull of all points{\small{
\begin{align}
&\Bigl\{ (R_1,R_2,R_3) \in \posreals^{3}:
\Bigr. \nonumber \\
&R_i < \min  \Bigl( \D_1 I(\Xa{i}{1};\Yb{j}{1} \vert
\Xa{k}{1}), \D_1 I(\Xa{i}{1};\Yb{k}{1} \vert \Xa{j}{1}), \Bigr. \nonumber \\
&\qquad \qquad \qquad \quad \Bigl. \D_1 I(\Xa{i}{1};\Yr \vert \Xa{j}{1},\Xa{k}{1}), \D_2 I(\Xr;\Yb{i}{2})\Bigr),\nonumber\\
&R_i + R_j < \min \Bigl(\D_1 I(\Xa{i}{1},\Xa{j}{1};\Yb{k}{1}), \Bigr. \nonumber\\
& \Bigl. \qquad \qquad \qquad \qquad \quad \D_1 I(\Xa{i}{1}, \Xa{j}{1};\Yr \vert \Xa{k}{1}) \Bigl); (i,j,k) \in {\mathbb{P}_{3}},\nonumber \\
& \Bigl. R_1 + R_2 + R_3 <  \D_1
I(\Xa{1}{1},\Xa{2}{1},\Xa{3}{1};\Yr) \Bigr\}, \nonumber
\end{align}}}
\noindent over all joint distributions $p^{(1)}(x_{a_1})p^{(1)}(x_{a_2})p^{(1)}(x_{a_3})p^{(2)}(x_r)$, over the alphabet
${\mathcal{X}_{a_1}} \times {\mathcal{X}_{a_2}} \times {\mathcal{X}_{a_3}} \times {\mathcal{X}_{r}}$.

\begin{proof1} {\em Encoding}: For $i=1,2,3$, the transmitters generate $2^{nR_{i}} $ random $n\D_1$-length sequences $\bXa{i}{1}$
according to the distributions $p^{(i)}(x_{a_i})$ respectively to construct their codebooks. Then, $\{a_i\}_{i=1}^{3}$ pick
their messages ${\{w_{a_i}\}}_{i=1}^{3}$ at random from the index sets ${\{\left[1,2^{nR_i}\right]\}}_{i=1}^{3}$, respectively,
where $[1,M] \eqdef \left\{1,2,\cdots,M\right\}$, and send ${\{\bXa{i}{1}(w_{a_{i}})\}}_{i=1}^{3}$ respectively. The relay node
also generates $2^{n{\sum_{i=1}^{3}R_{i}}}$ random $n\D_2$-length sequences $\bXr(w_{a_1},w_{a_2},w_{a_3})$. The relay node $r$
estimates $\tilde{w}_{a_1}$, $\tilde{w}_{a_2}$, and $\tilde{w}_{a_3}$ at the end of phase 1 and then broadcasts the signal
$\bXr(\tilde{w}_{a_1},\tilde{w}_{a_2},\tilde{w}_{a_3})$.

{\em Decoding and error analysis}: The decoding at the receiver nodes ${\{b_i\}}_{i=1}^{3}$ is done in two steps. Each receiver
node (e.g., $b_1$) decodes the messages of the other two pairs (e.g., ${w_{a_2}}$ and ${w_{a_3}}$) at the end of the first
phase, using the side information received during the first phase. Second, using this information, the receiver node tries to
find its own desired message (e.g., ${w_{a_1}}$), after the second phase. The detailed decoding process and error analysis at
nodes $r$ and $b_1$ is further explained in the sequel. Note that the analysis is similar for nodes $b_2$ and $b_3$, due to the
symmetry of the network.

{\em Relay}: The relay decodes ${\tilde{w}_{a_1}}$, ${\tilde{w}_{a_2}}$, and ${\tilde{w}_{a_3}}$ at the end of phase 1, using
joint typical decoding, if this triple is the only one satisfying $(\bXa{1}{1}(\tilde{w}_{a_1}), \bXa{2}{1}(\tilde{w}_{a_2}),
\bXa{3}{1}(\tilde{w}_{a_3}), \bYr ) \in {{T}}_{\epsilon}^{(1)}$. Indeed, during phase 1 with a block length of $n\D_1$, a MAC
is formed from $a_1$, $a_2$, and $a_3$ to $r$. The error analysis of the MAC is known \cite{Cover:2006} that we will have
${\sf{Pr}}\left(\tilde{w}_{a_1} \neq w_{a_1}\right), {\sf{Pr}}\left(\tilde{w}_{a_2} \neq {w}_{a_2} \right),
{\sf{Pr}}\left(\tilde{w}_{a_3} \neq {w}_{a_3} \right) \rightarrow 0$, as $n \rightarrow \infty$ if
\begin{align}
R_{S} & < \D_1 I({X_{S}^{(1)}};\Yr \vert {X_{S^c}^{(1)}}), \label{relay_DF}
\end{align}
\noindent for all $S \subseteq \{1,2,3\}$, where $S^c = \{1,2,3\} - S$.

{\em Receiver $b_1$}: Terminal node $b_1$, decodes ${\hat{w}_{a_2}}$ and ${\hat{w}_{a_3}}$ after phase 1 from the received
signal $\bYb{1}{1}$, if there exists a unique pair ${\hat{w}_{a_2}}$ and ${\hat{w}_{a_3}}$, such that
$(\bXa{2}{1}(\hat{w}_{a_2}), \bXa{3}{1}(\hat{w}_{a_3}), \bYb{1}{1}) \in {{T}_{\epsilon}}^{(1)}$. Using the error analysis of
the MAC \cite{Cover:2006}, we have ${\sf{Pr}}\left(\hat{w}_{a_2} \neq {w}_{a_2} \right), {\sf{Pr}}\left(\hat{w}_{a_3} \neq
{w}_{a_3} \right) \rightarrow 0$, if 
\begin{align}
R_2 < &  \D_1  I(\Xa{2}{1};\Yb{1}{1} \vert \Xa{3}{1})
, \label{b1_DF1}\\
R_3 < &  \D_1 I(\Xa{3}{1};\Yb{1}{1} \vert \Xa{2}{1}),  \label{b1_DF2} \\
R_2 + R_3 & < \D_1  I(\Xa{2}{1},\Xa{3}{1};\Yb{1}{1} ).
\label{b1_DF3}
\end{align}
Finally, the receiver $b_1$ estimates its desired message $w_{a_1}$, by looking for a unique $\hat{w}_{a_1}$ such that
$(\bXr(\hat{w}_{a_1},\hat{w}_{a_2},\hat{w}_{a_3}), \bYb{1}{2}) \in {{T}_{\epsilon}}^{(1)}$. Therefore the error event
$E_1^{(2)} = \left\{ w_{a_1} \neq \hat{w}_{a_1} \right\}$ can be written as
\begin{equation}
E_1^{(2)} = E_1^{(2)} \cap \left[ \left( \bar{E}_{r} \cap
\bar{E}_{s,1} \right) \cup \left( {E}_{r} \cup {E}_{s,1} \right)
\right],\nonumber
\end{equation}
\noindent where ${E}_{r} = {E}_{1,r}^{(1)} \cup {E}_{2,r}^{(1)} \cup {E}_{3,r}^{(1)}$ and ${E}_{s,1} = {E}_{2,1}^{(1)} \cup
{E}_{3,1}^{(1)}$ are the events of decoding error at relay and $b_1$ after phase 1, respectively. Consequently, by the AEP
property, the probability of the error event $E_1^{(2)}$ can be upper bounded by
\begin{align}
{\sf{Pr}}\bigl[E_1^{(2)}\bigr] \leq & {\sf{Pr}}\bigl[ E_1^{(2)}
\bigl\vert  \bar{E}_{r} \cap \bar{E}_{s,1} \bigr] + {\sf{Pr}}\bigl[
{E}_{r} \cup {E}_{s,1} \bigr] \nonumber\\
\leq  2^{{-n}R_1} & 2^{n\D_2 \left( I(\Xr;\Yb{1}{2}) -2\epsilon
\right)} + {\sf{Pr}}\bigl[ {E}_{r} \cup {E}_{s,1}
\bigr]. 
\label{error_DF_final}
\end{align}
By choosing $R_1 < \D_2 ( I(\Xr;\Yb{1}{2}) -2\epsilon )$, and applying all inequalities
(\ref{relay_DF}), 
the right hand side of {(\ref{error_DF_final})} will vanish as $n \rightarrow \infty$. By following the same argument for
decoding at nodes $b_2$ and $b_3$, we can derive two similar set of inequalities to {(\ref{b1_DF1})--(\ref{b1_DF3})} and also
conclude that $R_2 < \D_2 ( I(\Xr;\Yb{2}{2})-2\epsilon )$ and $R_3 < \D_2 ( I(\Xr;\Yb{1}{3}) -2\epsilon )$. Finally, since
$\epsilon > 0$ is arbitrary, the conditions of Theorem \ref{Theorem:DF} hold.
\end{proof1}
\end{theorem}
{\em Gaussian Case}: For the case, where all the channels are AWGN,
and we have $R=R_1=R_2=R_3$, it can be easily seen that the
conditions of Theorem \ref{Theorem:DF}, will reduce to
\begin{align}
R & < \min (\D_1,\D_2) \log(1+\snr), \label{df_gaussian1}\\
2R & < \D_1 \log(1+2\snr),   \\
3R & < \D_1 \log(1+3\snr). \label{df_gaussian2}
\end{align}
\subsection{Amplify-and-Forward (AF)} \label{sec:notation:defs}
In this section, we consider a two phase amplify-and-forward (AF) strategy, in which the relay simply forwards a scaled version
of the signal it receives after phase 1, imposing the equality $\D_1 = \D_2 ={1/2}$. Thus, we assume continuous input and
output alphabets for the terminals and the relay, which is inherent to an AF strategy.

\begin{theorem}
\label{Theorem:AF} An achievable region for the star relay network with the two phase AF MBC protocol is the closure of the
convex hull of all points{\small\begin{align} &\Bigl\{ (R_1,R_2,R_3) \in \posreals^{3}: \ \Biggr. R_i < {1 \over 2}
I(\Xa{i}{1};\Yb{i}{2} \vert
\Xa{j}{1},\Xa{k}{1}), \label{AF_Region} \\
&R_i + R_j < \min \Bigl({1 \over 2}I(\Xa{i}{1},\Xa{j}{1};\Yb{i}{1},\Yb{i}{2} \vert \Xa{k}{1}), \Bigr. \nonumber\\
&\qquad \qquad \qquad \Bigl. {1\over 2} I(\Xa{j}{1}, \Xa{i}{1}; \Yb{j}{1},\Yb{j}{2} \vert \Xa{k}{1}) \Bigl); \quad (i,j,k) \in {\mathbb{P}_{3}} \nonumber  \\
&\Bigr. R_1 + R_2 + R_3 < \min_{i\in\{1,2,3\}} \Bigl({1 \over 2} I(\Xa{1}{1},\Xa{2}{1},\Xa{3}{1};\Yb{i}{1},\Yb{i}{2}) \Bigr) \ \Bigr\},\nonumber
\end{align}}
\noindent over all joint distributions $p^{(1)}(x_{a_1})p^{(1)}(x_{a_2})p^{(1)}(x_{a_3})$, over the alphabet
${\mathcal{X}_{a_1}} \times {\mathcal{X}_{a_2}} \times {\mathcal{X}_{a_3}}$.

\begin{proof1}
{\em Encoding}: For $i=1,2,3$, the transmitter nodes $a_i$ generate $2^{nR_{i}}$ random ${n \over 2}$-length sequences
$\bXa{i}{1}$, according to the distributions $p^{(1)}(x_{a_i})$  and broadcast the sequences that correspond to the messages
$w_{a_{i}}$ during phase 1. At the end of phase 1, the relay scales (or amplifies) its received signal by a coefficient
$\alpha$ and generates the ${n \over 2}$-length sequence $\bXr = \alpha \bYr$. The relay then broadcasts $\bXr$ to the side
nodes.

{\em Decoding and error analysis}: The decoding at receiver nodes ${\{b_i\}}_{i=1}^{3}$ is done after phase 2. After phase 1,
each receiver node (e.g., $b_1$) buffers the sum signal it has received from its neighbor nodes (e.g.,$\bYb{1}{1}$). It uses
this side information in conjunction with the signal it receives from the relay after phase 2 (e.g., $\bYb{1}{2}$), to perform
a jointly typical decoding and decode its own
desired message (e.g., ${w_{a_1}}$). 
Decoding and error analysis at $b_1$ is explained in the sequel. Because of the symmetry, the analysis is similar for nodes
$b_2$ and $b_3$.

{\em Receiver $b_1$}: After phase 2, terminal node $b_1$, looks for the triples ${\hat{w}_{a_1}}$, ${\hat{w}_{a_2}}$, and
${\hat{w}_{a_3}}$, for which $( \bXa{1}{1}(\hat{w}_{a_1}), \bXa{2}{1}(\hat{w}_{a_2}), \bXa{3}{1}(\hat{w}_{a_3}), \bYb{1}{1},
\bYb{1}{2} ) \in T_{\epsilon}^{(2)}$. It then claims the index ${\hat{w}_{a_1}}$, as the decoded message, if the first element
of all such triples is identically ${\hat{w}_{a_1}}$. The error event $E_{1}^{(2)}$ is thus the union of four events
\begin{align}
\label{union_error_AF} E_{1}^{(2)} = E_{\emptyset,1}^{(2)} \cup E_{\{2\},1}^{(2)} \cup
E_{\{3\},1}^{(2)} \cup E_{\{2,3\},1}^{(2)},
\end{align}
\noindent where we define
\begin{align}
&\hspace{-1.8cm} E_{{S},i}^{(2)} = \bigcup_{\substack{{\hat{w}_{a_{i}}} \neq {w_{a_i}} \\ {\hat{w}_{a_{\ell}}} \neq {w_{a_\ell}} : \ell \in S }}
\bigl( \bigr. \bXa{i}{1}(\hat{w}_{a_i}), \{\bXa{\ell}{1}({w}_{a_\ell})\}_{\ell \in S^{c}},\nonumber
\end{align}\vspace{-1cm}
\begin{align}
& \qquad \qquad \qquad \qquad \quad \{\bXa{\ell}{1}(\hat{w}_{a_\ell})\}_{\ell \in S},
\bYb{i}{1}, \bYb{i}{2} \bigl. \bigr) \in T_{\epsilon}^{(2)}, \nonumber
\end{align}
\noindent for brevity, where $S^{c} = \{1,2,3\} - \{i\} -S$. From \eqref{union_error_AF} and by applying the union bound and
the AEP property, we have
\begin{align}
{\sf{Pr}} \bigl[E_1^{(2)}\bigr] & \leq  2^{nR_1}2^{{-n \over 2} \left( I
(\Xa{1}{1};\Yb{1}{2} \vert
\Xa{2}{1},\Xa{3}{1}) - 2\epsilon \right)} \nonumber \\
&\hspace{-1.2cm} + 2^{n(R_1+R_2)}2^{{-n \over 2} \left(I
(\Xa{1}{1},\Xa{2}{1};\Yb{1}{1},\Yb{1}{2} \vert \Xa{3}{1})-3\epsilon\right)} \nonumber\\
& \hspace{-1.2cm}+ 2^{n(R_1+R_3)}2^{{-n \over 2}\left( I
(\Xa{1}{1},\Xa{3}{1};\Yb{1}{1},\Yb{1}{2} \vert \Xa{2}{1}) -3\epsilon \right)} \nonumber\\
& \hspace{-1.2cm} + 2^{n(R_1+R_2+R_3)}2^{{-n \over 2} \left( I
(\Xa{1}{1},\Xa{2}{1},\Xa{3}{1};\Yb{1}{1},\Yb{1}{2}) -4\epsilon \right)}. \nonumber
\end{align}

By following the same argument for decoding at nodes $b_2$ and $b_3$, we can derive similar upper bounds on the corresponding
error probabilities $ {\sf{Pr}}\bigl[E_2^{(2)}\bigr]$ and ${\sf{Pr}}\bigl[E_3^{(2)}\bigr]$. Since $\epsilon
> 0$, is chosen arbitrarily, by choosing $R_1$, $R_2$, and $R_3$
according to the conditions of Theorem \ref{Theorem:AF}, we can upper bound the error probabilities by $0$ as $n \rightarrow
\infty$.
\end{proof1}

{\em Gaussian case}: Now, we assume the channels to be
AWGN. 
To maximize the mutual information measures, all of the nodes employ Gaussian complex codebooks with transmit power $P$ for
both
phases. 
We consider the exchange rate and rewrite the conditions of Theorem \ref{Theorem:AF}, for AWGN channels. The result is stated
as a corollary.
\begin{corollary}\label{AF_fading}
An achievable exchange rate of the star network with the two phase AF protocol and AWGN channels is given by $R <
\underline{\mathcal{C}}_{\sf AF}$, where
\begin{align}
&  \underline{\mathcal{C}}_{\sf AF} = \min \Biggl\{ {1\over 2}  \log   \left[1+\left({\snr \over 1 +
4\snr}\right) \snr\right], \Biggr. \nonumber\\
& \quad {1\over 4} \log  \biggl[1+ \biggl({1 +6\snr \over 1 + 4\snr}\biggr)
 \snr +
\left({\snr \over 1 + 4\snr}\right) \snr^2 \biggr], \nonumber \\
&\quad \Biggl.  {1\over 6} \log \biggl[1+ \biggl({2 + 11\snr \over 1 +
4\snr}\biggr) \snr + \left({2\snr \over 1 + 4\snr}\right) \snr^2
\biggr] \Biggr\}. \label{AF_ex_rate}
\end{align}

\begin{proof}
This follows by (\ref{Gaussian1})--(\ref{Gaussian3}), Theorem \ref{Theorem:AF} and $\Xr = \alpha \Yr$, where $\alpha^2 ={\snr
\over 1 + 3\snr}$, due to the node power constraint, $P = E{\lvert\Xr\rvert ^ 2} = {\alpha^2} E{\vert\Yr\vert ^ 2}$. 
\end{proof}
\end{corollary}
\end{theorem}\vspace{-.6cm}
\subsection{Lattice Coding}
\begin{theorem}
\label{Theorem:Lattice} An inner bound to the exchange capacity of the star relay network with the MBC protocol and
synchronized AWGN channels is given by
\begin{align}\label{lattice_ex}
R < \D_1 \log ({1 / 3} + \snr),
\end{align}
provided that $\D_1 \log ({1 / 3} + \snr) < \D_2 \log ({1} + \snr)$. By maximizing the right hand side of (\ref{lattice_ex})
over $\D_1$, subject to this condition, we find every $R < \underline{\cal{C}}_{\sf Latt.}$ is achievable where
\begin{align}
\label{lattice_ex_rate}
\underline{\cal{C}}_{\sf Latt.} = \frac{\log (1+ \snr) \log ({1 \over 3} +
\snr)} {\log ({1} + \snr) + \log ({1 \over 3} + \snr)}.
\end{align}
\begin{proof1}
The idea of the proof is an extension of the one in \cite{Narayanan:2007}, which uses nested lattices to encode and decode the
information. The difference is that we also exploit the side information available to the terminals after phase 1. In
particular, we consider the star relay network with perfectly synchronized AWGN channels. We use a fine lattice, nested in a
coarse lattice with second moment $P$, with its points located in the basic Voronoi region of the coarse lattice as the
codewords. We denote the coarse lattice by $\cal{L}$ and the fine lattice nested in it by ${\cal{L}}_f$, thus ${\cal{L}}
\subseteq {\mathcal{L}_{f}}$. The basic Voronoi region of lattices are denoted by $S(\mathcal{L})$ and $S({\mathcal{L}_{f}})$.
See \cite{Narayanan:2007} for more details on lattices.

{\em Encoding}: The encoders ${\{a_i\}}_{i=1}^{3}$ map their messages $\{{w_{a_i}}\}_{i=1}^{3}$ on to the $n\D_1$-dimensional
points ${\{\bc_{i}({w_{a_i}})\}}_{i=1}^{3}$ of the basic Voronoi cell ${{\mathcal{L}}_{f}} \cap S(\mathcal{L})$. 
The encoders then generate random $n\D_1$-length {\em dither} vectors ${\{\bd_{i}\}}_{i=1}^{3}$ according to the uniform
distribution over $S(\cal{L})$. The dither vectors are mutually independent and known to the relay and receivers. The transmit
signals $\bXa{i}{1}(w_{a_{i}})$ are then constructed according to the following rule
\begin{align}
\label{lattice_encoding} \bXa{i}{1} & = \bc_{i} - \bd_{i} \mod
{\mathcal{L}}, \quad i = 1,2,3.
\end{align}

{\em Decoding}: The relay decodes $\bc_{r} = ( \bc_{1} + \bc_{2} + \bc_{3} ) \mod \mathcal{L}$, after phase 1 and then
transmits the index of $\bc_{r}$, using random coding. The receivers (e.g., $b_1$) decode the modular sum of the other pairs
codewords (e.g., $\bc_{2} + \bc_{3} \mod \mathcal{L}$), after phase 1. Finally the receivers decode $\bc_{r}$ and find the
desired codeword by a modular subtraction. The details of decoding at both phases are as follows.

1) {\em Phase 1 Decoding}: Since the channels are AWGN and synchronized, after phase 1, the relay receives the sum of signals
transmitted by ${\{a_i\}}_{i=1}^{3}$ and the receivers ${\{b_i\}}_{i=1}^{3}$ receive the sum of signal pairs broadcasted by
their neighbor nodes. The relay and terminals perform similar lattice decoding procedures after phase 1. The received vector at
the relay after phase 1 is
\begin{align}
\label{relay_lattice} \bYr & = \bXa{1}{1} + \bXa{2}{1} + \bXa{3}{1}
+ \bZr,
\end{align}
\noindent from which the relay decodes $\bc_{r}$, which is itself a codeword, due to the group property of the set
${{\mathcal{L}}_{f}} \cap S(\mathcal{L})$ under addition $\mod {\mathcal{L}}$. For this purpose, the relay forms the signal
$\hat{\bc}_{r} = (\gamma \bYr + \bd_1 + \bd_2 + \bd_3 ) \mod {\mathcal{L}}$, where $\gamma \in \mathbb{R}$, is a scaling
coefficient, determined to maximize the achievable exchange rate (i.e., maximize the number of fine lattice points in the basic
cell of the coarse lattice). From (\ref{lattice_encoding}) and (\ref{relay_lattice}),
\begin{align}
\label{lattice_decoding}  {\hat \bc_{r}} & = \bigl[
(\gamma (\bXa{1}{1} + \bXa{2}{1} + \bXa{3}{1} +
\bZr)+ \sum_{i=1}^{3}\bd_{i})\bigr] \mod {\cal{L}}, \nonumber \\
& = \bigl[\bc_{r} + \gamma \bZr - (1-\gamma)\sum_{i=1}^{3}\bXa{i}{1}
\bigr] \mod {\mathcal{L}}.
\end{align}
Because ${\{\bc_{i}\}}_{i=1}^{3}$ are independent of the noise and ${\{\bXa{i}{1}\}}_{i=1}^{3}$, we can rewrite
(\ref{lattice_decoding}) as $\hat{\bc}_{r} = \bc_{r} + {\tilde{{\bf{Z}}}_{r}^{(1)}}$, where ${\tilde{{\bf{Z}}}_{r}^{(1)}} =
(\gamma \bZr - (1-\gamma)\sum_{i=1}^{3}\bXa{i}{1}) \mod {\mathcal{L}}$, is an equivalent noise term added to the desired signal
with power $\tilde{N} = {\gamma^2}N + 3P(1-\gamma)^{2}$. The optimal value of $\gamma$ to minimize the equivalent noise power
is $\gamma^{*} = {3P \over {3P + N}}$, and the corresponding optimal noise power is $\tilde{N}^{*}_{r} = {3PN \over {3P + N}}$.
By the properties of lattices and error probability of lattice codes for AWGN channel \cite{Erez_Litsyn_Zamir:2005},
\cite{Erez_Zamir:2004}, the second moment of the fine lattice should be chosen $D({\mathcal{L}_{f}}) = \tilde{N}^{*}_{r} +
\delta$, for arbitrary $\delta>0$, in order to have ${\pr}\left[\hat{\bc}_{r} \neq \bc_{r}\right] \rightarrow 0$, as $n
\rightarrow 0$. By letting $\delta \rightarrow 0$, the maximum packed number of fine lattice points in the basic cell is
$\lvert {{\mathcal{L}}_{f}} \cap S(\mathcal{L}) \rvert = \left[{D({\mathcal{L}}) / D\left({{\mathcal{L}}_{f}}\right)
}\right]^{n\D_{1}} = {\left({1/3}+\snr\right)}^{n\D_{1}}$. Since the number of codewords $2^{nR}$ is at most the number of
basic cell points 
for reliable MAC decoding at the relay as $n \rightarrow \infty$, we require $R < \D_1\log({1 / 3} + \snr)$.
The phase 1 decoding at the lateral receivers ${\{b_{i}\}_{i=1}^{3}}$ follows the same outline as the relay. Noting that these
nodes should decode the sum of two codewords$\mod {\mathcal{L}}$, rather than three, we obtain the condition $R < \D_1\log({1 /
2} + \snr) $ to guarantee correct decoding at lateral terminals as $n\rightarrow \infty$.

2) {\em BC Decoding}: The relay uses a random code to convey $\hat\bc_{r}$ to the nodes $\{b_{i}\}_{i=1}^{3}$ in $n\D_2$
channel uses.
The decoded signal of $\hat\bc_{r}$ at each $b_i$ is referred to as ${\hat{\hat{\bc}}}_{r,i}$. The relay codebook consists of
$\lvert {{\mathcal{L}}_{f}} \cap S(\mathcal{L}) \rvert$ codewords. Therefore, from the classical AWGN channel error analysis,
$\pr{\bigl[{\hat{\hat{\bc}}}_{r,i} \neq \hat\bc_{r}\bigr]} \rightarrow 0$, as $n \rightarrow \infty$, as long as $ \D_1
\log(1/3+\snr) < \D_2 \log(1+ \snr)$. The proof is complete by noting that, by the union bound, the decoding error probability
at  each $\{b_{i}\}_{i=1}^{3}$ will vanish if both MAC and BC phases decodings are separately asymptotically reliable.
\end{proof1}
\end{theorem}
\section{Comparison and Numerical Results}
{\begin{figure} \label{performance} {\hspace{0.15cm}{\resizebox*{3.35in}{2.1in}{\includegraphics{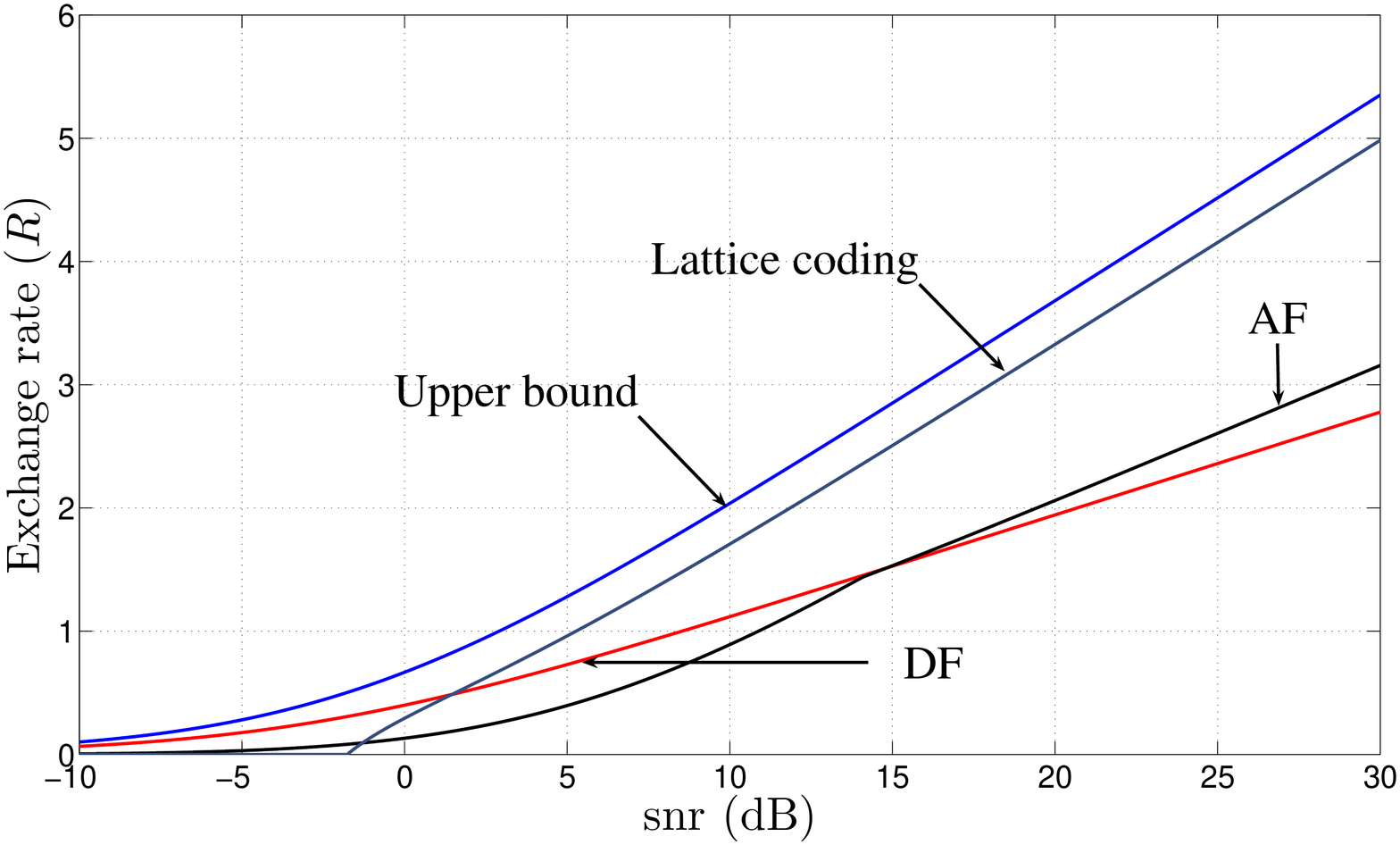}}
\center{\caption{Achievable exchange rates for different strategies and upper bound.}}}} \vspace{-0.4cm}
\end{figure}}
In this section, we compare the performance of three schemes examined in this paper (DF, AF, and lattice), for the two phase
protocol with AWGN channels, in terms of the exchange rate and their gap to the upper bound. 
From (\ref{df_gaussian1})--\eqref{df_gaussian2}, we find the optimized over $\D_1$ achievable rate for DF as a function of
$\snr$ as
\begin{equation}\label{ultimate_DF}\underline{\mathcal{C}}_{\sf DF}= \frac{\log (1+ \snr) \log ({1} + 3\snr)} {3\log ({1} +
\snr) + \log ({1} + 3\snr)}.\end{equation}

We also find the global upper bound by optimizing (\ref{upper1}) over $\D_1$ to
be\begin{equation}\label{ultimate_UB}\overline{\mathcal{C}}_{\sf UB}= \frac{\log (1+ \snr) \log ({1} + 3\snr)} {\log ({1} +
\snr) + \log ({1} + 3\snr)}.\end{equation}

\Fig 2 illustrates the achievable curves \eqref{ultimate_DF}, \eqref{AF_ex_rate}, and \eqref{lattice_ex_rate} for DF, AF, and
lattice-based schemes respectively and \eqref{ultimate_UB} for the upper bound. It can be seen that the lattice-based strategy
outperforms other schemes for high $\snr$ values, where it is asymptotically within $\log(3)/4 \simeq 0.4$ bit of the upper
bound, while at the worst case, it has a gap of $\log(3)\log(5/3)/\log(5) \simeq 0.5$ bit. If we plot the best of the curves in
terms of $\snr$, the compound scheme will be within $0.34$ bit of the upper bound for all $\snr$ values.

\section{Conclusion}
We considered a wireless network with three source-sink pairs of terminals that want to communicate with the help of a relay,
using a two phase protocol. We derived an upper bound and three achievable exchange rates. As a main part, we proposed codes
for the system based on high-dimensional lattices and incorporated relaying, as well as joint physical and network layer coding
with the use of side information. We showed that the lattice coding scheme can achieve an exchange rate within $0.5$ bit of the
upper bound.
\bibliographystyle{IEEEtranS}
\bibliography{main2}

\end{document}